\documentclass[fleqn,12pt,letterpaper]{article}
\usepackage{amsfonts}
\usepackage{amssymb}
\usepackage{amsmath}
\usepackage{amsthm}
\usepackage{subcaption}
\usepackage{adjustbox}
\usepackage{booktabs}
\usepackage{xcolor}

\usepackage{bbm}
\usepackage{enumitem}
\usepackage{tabularx}
\usepackage{multirow}
\usepackage{float}
\usepackage{comment}
\usepackage{graphicx}
\usepackage{epstopdf}
\usepackage{color}
\usepackage[margin=2.57cm]{geometry}
\usepackage{threeparttable}
\usepackage{siunitx}
\definecolor{darkred}{rgb}{0.5,0.2,0.2}
\usepackage[pdftitle={RV CL},pdfauthor={Mikkel Bennedsen},colorlinks=TRUE,allcolors=darkred]{hyperref}
\usepackage{booktabs}
\usepackage{pdflscape}
\usepackage[round]{natbib}
\usepackage{subcaption}
\usepackage{mwe}
\usepackage{adjustbox}
\newcommand{\br}{\bigr)}
\newcommand{\bl}{\bigl(}
\usepackage{dsfont}
\usepackage{cleveref}
\usepackage{float}
\floatstyle{plaintop}
\restylefloat{table}

\newcommand{\interior}[1]{%
  {\kern0pt#1}^{\mathrm{o}}%
}

\restylefloat{table}

\theoremstyle{plain}
\newtheorem{theorem}{Theorem}[section]

\newtheorem{assumption}{Assumption}[section]

\newtheorem{lemma}{Lemma}[section]
\newtheorem{proposition}{Proposition}[section]

\theoremstyle{definition}

\theoremstyle{remark}
\newtheorem{remark}{Remark}[section]

\def\bi{\begin{itemize}}
\def\ei{\end{itemize}}

\allowdisplaybreaks
\graphicspath{{Figures/}}
\numberwithin{equation}{section}

\newif\ifi

\usepackage{setspace}

\usepackage{caption}

\usepackage{etoolbox}
\makeatletter
\patchcmd{\H@@footnotetext}{\reset@font\footnotesize}{\reset@font\singlespacing\footnotesize}{}{}
\patchcmd{\@footnotetext}{\reset@font\footnotesize}{\reset@font\singlespacing\footnotesize}{}{}
\makeatother

\begin{document}

\title{Microstructural Foundations of Rough Noise\footnote{We are grateful to Carsten Chong for fruitful discussions on rough noise. We benefitted
from comments by and discussions with Kim Christensen, Mikko Pakkanen, Bent Jesper Christensen, and Mikkel Bennedsen, plus seminar and conference participants at University
of Macau, the 4th Italian Meeting on Probability and Mathematical Statistics in Rome, the SIAM Conference on
Financial Mathematics and Engineering 2025 in Miami, and the 2023 annual SoFiE conference in Seoul, South Korea.}}

\author{Peter Korsbakke Christensen\thanks{
Department of Economics and Business Economics and CoRE, 
Aarhus University, 
Fuglesangs All\'e 4,
8210 Aarhus V, Denmark.
E-mail:\
\href{mailto:pchr@econ.au.dk}{\nolinkurl{pchr@econ.au.dk}}. 
}
\and
Anders Norlyk\thanks{
Department of Economics and Business Economics and CoRE, 
Aarhus University, 
Fuglesangs All\'e 4,
8210 Aarhus V, Denmark.
E-mail:\
\href{mailto:amn@econ.au.dk}{\nolinkurl{amn@econ.au.dk}}. 
}
}

\maketitle
\begin{abstract}
\singlespacing
Recently, it has been proposed to model the microstructure noise in prices by a continuous-time process with continuous sample paths that are rougher than those of a standard Brownian motion. In this paper, we propose a microstructural model for the tick-by-tick price changes that explicitly separates the permanent price changes from the fleeting price changes due to noise. We show how this model converges to a standard semimartingale model for the permanent price process, plus a rough noise term originating from the fleeting price changes on the macro scale. This provides a microstructural foundation for the rough-noise model. We then develop a GMM estimation method applicable to tick-by-tick data, together with a formal test for rough noise. We show that the estimator and test work in finite samples through a simulation study, and apply them to tick-by-tick data on Dow Jones Industrial Average constituents in 2024. Because our estimator is designed for tick-by-tick data, we estimate roughness at the daily level, revealing substantial day-to-day variation. We find that rough noise, while present, is not universal: even when detected, the roughness index is often mild, and it is most pronounced on days dominated by short-run price reversals.
\end{abstract}


\newpage
\doublespacing
\section{Introduction}

It is a well-known stylized fact in financial econometrics that prices observed
at high frequencies are contaminated by microstructure noise, stemming from
bid--ask bounces, transaction costs, rounding, and related frictions
\citep{Hasbrouck2007,JacodEtAl2017}. The observed logarithmic price is typically
decomposed as $Y_t = X_t + Z_t$, where the efficient price $X$ is a continuous
It\^o semimartingale and $Z$ is additive noise \citep{AitSahaliaEtAl2005}. When
prices are observed on a grid $\{i\Delta_n : i=1,\ldots,[T/\Delta_n]\}$, the
noise is conventionally modeled as a discrete-time sequence
$Z_{i\Delta_n}=\varepsilon_i$, specified, for example, as white noise
(\citealp{ChristensenEtAl2010,preAvg09}), as an AR- or MA-type series
(\citealp{PeterAsgerRV06,SahaliaMyklandZhang11}), or as arising from rounding
(\citealp{DelattreJacod1997,LiMykland2010}). The noise also has well-documented
consequences for inference. In particular, realized variance, which
consistently estimates integrated volatility in the absence of noise
\citep{andersen_etal01a}, diverges as the sampling frequency increases
\citep{AndersenEtAl2000}.

Recently, \citet{ChongEtAl2021} argued that discrete noise sequences are
incompatible with two empirical features of high-frequency data. First,
realized variance diverges at a rate slower than the $\Delta_n^{-1}$ implied by
stationary discrete noise. Second, as $\Delta_n \to 0$, increments of noise shrink, whereas the noise component remains non-shrinking, a feature discrete time noise cannot reproduce.
 They therefore propose modeling $Z$
as a continuous-time process with sample paths \emph{rougher} than those of
Brownian motion, which accommodates both features. The roughness of the noise
matters for inference, as classical noise-robust techniques such as
preaveraging \citep{preAvg09} are no longer consistent estimators of integrated
volatility under rough noise \citep{ChongEtAl2021}.

While the rough-noise model has appealing properties on the macroscopic scale,
an important question is whether it is consistent with a realistic
microstructural description of prices. That is, does there exist a tick-by-tick
model, replicating the salient features of ultrahigh-frequency prices, whose
scaling limit is of the form $Y_t=X_t+Z_t$ with $Z$ a continuous-time rough
process? Providing such a microstructural foundation is the aim of the present
paper.

Our first contribution is a microstructural foundation for rough noise.
Building on the model of fleeting price moves of \citet{Shephard2017}, we model
tick-by-tick prices on an integer grid, decomposed into a permanent and a
transitory component. The permanent component is an integer-valued L\'evy
process, the continuous-time analogue of a random walk on the tick grid, while
the transitory component is a trawl process capturing short-lived price
distortions and their subsequent correction. In an asymptotic framework
inspired by \citet{Rosenbaum2018}, we show that the rescaled price process
converges to a Bachelier-type semimartingale plus a Gaussian moving average
whose kernel is determined explicitly by the trawl function. By choosing the
trawl function appropriately, the limiting noise attains any roughness index
$\alpha\in(-1/2,0]$, matching the local behavior, and hence the roughness, of
the noise specification of \citet{ChongEtAl2021}. Moreover, the mechanism is transparent, as roughness arises when distortions
of very short duration dominate, being created and corrected almost
instantaneously.

Our second contribution is an inference methodology applicable directly on data at the highest frequency. We construct a GMM estimator of the model parameters from
second moments of price increments across multiple lags, and establish its
consistency and asymptotic distribution. Since the null hypothesis of non-rough
noise places the roughness index on the boundary of the parameter space,
standard asymptotics fail. Using results of
\citet{Andrews2002}, we show that the estimator remains $\sqrt{n}$-consistent
under the null, with a one-sided normal limit with a point mass at zero, and we
derive a feasible test of the null of no roughness against a rough alternative.
A simulation study shows that the estimator is approximately unbiased and that
the test has good size and power in samples corresponding to a single trading
day.

Our third contribution is empirical. We apply the methodology to tick-by-tick
transaction data on the Dow Jones Industrial Average constituents in 2024.
Because the estimator operates at the tick level, we obtain a separate
roughness estimate for each ticker-day, in contrast to existing evidence based
on multi-day rolling windows. The daily estimates reveal that rough noise is
present but episodic. About one fifth of ticker-days with detectable noise
reject the null of no roughness, the estimated roughness is typically mild, and
detections concentrate on days dominated by short-run price reversals rather
than being explained by broad illiquidity. The episodic nature of rough noise also reconciles our estimates with the
stronger, stable roughness reported by \citet{ChongEtAl2021}. Their estimator
is computed over multi-day rolling windows and is dominated by the roughest
day in each window, so occasional rough days can generate stable rough
estimates even when the typical day is not rough.

Our paper relates to a literature that builds tick-level models consistent
with semimartingale limits. \citet{AitSahaliaJacod2020} construct a model of transaction prices that remain
positive, move by at most one tick at a time, and exhibit rapid runs of
successive up- or down-ticks, and show that, after rescaling, such a model can
converge to an It\^o semimartingale with stochastic volatility and jumps. \citet{BacryEtAl2013b,Bacry2013} use Hawkes processes to
reproduce the divergence of realized variance and the Epps effect
\citep{Epps1979}. None of these models generate rough noise in the limit. Our
setup is instead analogous to the microstructural foundations of rough
volatility, in which \citet{Rosenbaum2018} obtain rough volatility as the
scaling limit of nearly unstable Hawkes models. In our model, roughness enters
through a different channel, namely the behavior of the trawl function near
zero, and it appears in the noise rather than in the volatility.

The remainder of the paper is organized as follows.
Section~\ref{sec:TickModelChoice} motivates the tick-by-tick specification,
combining microstructure arguments with empirical evidence from transaction
data. Section~\ref{sec:simple_model} introduces the model and establishes its
macroscopic limit. Section~\ref{sec:estimation} develops the GMM estimator, its
asymptotic theory, and the test for rough noise. Section~\ref{sec:sim} examines
finite-sample performance, Section~\ref{sec:empirical} contains the empirical
application, and Section~\ref{sec:conclusion} concludes.

\section{Motivating the tick-by-tick model}\label{sec:TickModelChoice}

In this section, we motivate the tick-by-tick model that serves as the
microscopic starting point of the paper. The observed price is decomposed into
an efficient price component and a microstructure noise component,
$Y_t = X_t + Z_t$ for $t\ge 0$. Our goal is to specify these two components in
a way that is consistent with the tick structure of transaction prices.

In the high-frequency literature, the efficient price is typically modeled at
the macroscopic scale as an It\^o semimartingale. At the tick level, however,
transaction prices live on a discrete grid, and a microscopic model should
respect this discreteness directly. Prices also evolve in continuous time, and
both components should reflect this. For the efficient price the
continuous-time formulation is standard, while for the noise it is motivated by
\citet{ChongEtAl2021}, as discussed in the introduction. A third requirement
concerns the economic role of the efficient price. Abstracting from time
variation in risk premia, an efficient-market view implies that its increments
behave approximately as martingale increments, whose classical discrete-time
benchmark is the random walk \citep{Campbell1998}. These three
requirements---discreteness, continuous time, and martingale-like
increments---are jointly met by an integer-valued L\'evy process, the
continuous-time analogue of a random walk on the tick grid, which we therefore
adopt for the permanent component.

The transitory component \(Z_t\) should capture temporary deviations of
transaction prices from the efficient price. The canonical mechanism is
bid--ask bounces. Transaction prices alternate between bid and ask quotes,
which mechanically generates reversals in observed price changes, and hence
negative autocorrelation in high-frequency returns, even when the efficient
price itself has martingale increments \citep{Roll1984}. Related mechanisms
whose price impact is transitory rather than informational, such as temporary
liquidity pressure and inventory control, have similar effects
(\citealp{Stoll1978,Hasbrouck2007,Ait2009}), and negative serial dependence in
high-frequency returns is a well-documented feature of transaction prices
(\citealp{SahaliaMyklandZhang11,JacodEtAl2017}).

\citet{ChongEtAl2021} emphasize a related empirical feature, namely that the
presence of microstructure noise depends on whether prices are constructed
from transactions on a single exchange or from all
exchanges.\footnote{This distinction is closely related to Rule P.3 in the
cleaning procedure of \citet{Barndorff2009}, which recommends constructing
prices from a single exchange rather than consolidating transactions across
exchanges.} Bid--ask bounces generates negative autocorrelation already at the
single-venue level, but it cannot explain why the dependence strengthens once
transactions are consolidated across exchanges. A natural candidate for the
additional dependence is cross-venue latency. If different venues incorporate
information with different delays, switching between them in the consolidated
tape can generate rapid reversals, even when each venue's own price has
uncorrelated increments. In a simple \citet{Hasbrouck1995} model with prices
on two exchanges, both following a random walk with one lagging the other,
consolidated returns exhibit negative first-order autocorrelation. The
calculations are detailed in online Appendix~D. In Appendix~F we further examine whether these effects are present in our data. We test the null of no autocorrelation at the first lag using prices from a single exchange, and compare the results with those obtained from consolidated prices across all exchanges. The central finding is that, for the all-exchange prices, the distribution of the test statistics shifts sharply to the left and the null is rejected on \(55\%\) of the ticker-days. This clearly indicates that consolidating transactions across venues introduces or amplifies short-lived reversals in observed prices.

To capture such short-lived distortions, we follow the
continuous-time model of fleeting discrete price moves of \citet{Shephard2017}
and model the noise as a trawl process, a stationary integer-valued process in
which each distortion appears, persists for a random duration, and is then
corrected. The distribution of these durations is governed by a deterministic
function, the trawl function, which thereby controls the dependence structure
of the noise.

\Cref{sec:simple_model} formalizes this discussion. Both components of the
observed price are generated by a single integer-valued L\'evy basis evaluated
on two different sets in space--time. Evaluating the basis on a growing
rectangle yields the permanent component, an integer-valued L\'evy process,
while evaluating it on a moving set shaped by the trawl function yields the
transitory component. The behavior of the trawl function near zero then
determines the properties of the noise in the macroscopic limit. When it is
unbounded at zero, distortions of arbitrarily short duration dominate, and the
accumulation of these microscopic reversals produces rough noise at the
macroscopic scale.

\section{From Ultra-High-Frequency to High-Frequency} \label{sec:simple_model}
In this section, we introduce the model of \citet{Shephard2017} in detail. In
view of the preceding discussion, this model will serve as our microscopic
description of price formation at ultra-high frequency. Our aim is to show
that, under a suitable scaling, this microscopic model gives rise to a
Bachelier-type price model with additive rough noise, belonging to the class
of mixed semimartingale models considered by \citet{ChongEtAl2021}.

The macroscopic target of this section is the mixed semimartingale model of
\citet{ChongEtAl2021}. There, the observed price is $Y_t = X_t + Z_t$, where
$X$ is an It\^o semimartingale and the noise component satisfies
$Z_t = Z_0+\int_0^t g(t-s)\rho_s\,dW_s$. Here $(\rho_t)_{t\geq 0}$ is adapted
and locally bounded, while the kernel $g$ is of the form
\begin{equation}\label{eq:chong_kernel}
g(t)
=
K_H^{-1}t^{H-1/2}+g_0(t),
\qquad H\in(0,1/2),
\end{equation}
where $g_0$ is smooth with $g_0(0)=0$, and $K_H$ is a constant depending on
$H$. The noise is therefore a Gaussian moving average whose sample paths are
rougher than those of Brownian motion, in the sense made precise in
\Cref{subsec:macro_limit}, its roughness index is $\alpha=H-1/2<0$.

This model does not respect the tick structure of markets and is therefore
not suitable at the highest frequencies. However, microstructure noise is
still modeled explicitly, so the setting is not noise-free. With this in
mind, we view it as a model for prices at high frequency, but not at
ultra-high frequency: price discreteness is no longer a dominant feature, but
microstructure noise remains present.

In the remainder of this section, we first introduce the microscopic price
model, then establish its macroscopic limit, and finally construct a trawl
function whose limit matches the local behavior, and hence the roughness, of
the target kernel \eqref{eq:chong_kernel}.

\subsection{Microscopic Price Model}\label{subsec:simple_model}
Let \(L\) be an integer-valued, time-homogeneous L\'evy basis on \(\mathcal S\times\mathbb R\), where \(\mathcal S\subseteq\mathbb R_+\). That is, \(L\) assigns an integer-valued random variable \(L(E)\) to each Borel set \(E\subseteq\mathcal S\times\mathbb R\) with finite Lebesgue measure, is additive over disjoint sets, and the resulting random variables are independent. We write \(L_1\) for the L\'evy seed of \(L\), which may be identified in distribution with \(L(E_0)\) for any Borel set \(E_0\subseteq\mathcal S\times\mathbb R\) with \(\operatorname{Leb}(E_0)=1\), and we denote its L\'evy measure by \(\nu\).

In our setup $[0,1]\subset \mathcal{S}$.\footnote{In the literature this is typically an equality. In what follows, however, we will need this space to be larger for some results, so we adopt a more general setup.}
Let \(b\in(0,1)\), and let \(d:[0,\infty)\to\mathcal S\) be a non-increasing function satisfying that $\int_0^\infty d(s) - b ds <\infty$. Define $A
    =
    \{(x,s):s\leq 0,\ b\leq x<d(-s)\}$,
and let
    $A_t=A+(0,t)
    =
    \{(x,s):s\leq t,\ b\leq x<d(t-s)\}$, with $t\geq 0$. Finally, define
    $B_t=[0,b)\times(0,t]$, for $t\geq 0$. We then set $C_t=A_t\cup B_t$
and define the microscopic price process by
\[
    P_t
    =
    V_0+L(C_t)
    =
    V_0+L(A_t)+L(B_t),
    \qquad t\geq 0.
\]
The term $L(B_t)$ is a Lévy process and captures permanent price changes: once an event enters $B_t$, it never leaves. We therefore interpret $L(B_t)$ as the efficient price component. By contrast, $L(A_t)$ is a trawl process. The set $A_t$ moves through the Lévy basis over time, so points enter and later leave $A_t$. Hence $L(A_t)$ captures temporary price changes and is interpreted as microstructure noise. The function $d$ controls the shape and overlap of the sets $A_t$ over time and therefore determines the dependence structure of this transitory component. We refer to it as the trawl function. Since $A_t$ and $B_t$ are disjoint for each $t\geq0$, and since $L$ is independently scattered, the two components are independent. Finally, the initial value $V_0$ captures permanent price changes prior to time zero.

As argued in \citet{Shephard2017}, this specification captures several key features of tick-by-tick prices: prices move on a discrete grid, evolve in continuous time, and may exhibit a large fraction of fleeting price changes. The parameter \(b\) controls the relative size of the permanent component: larger values of \(b\) imply that a larger share of price changes is permanent rather than fleeting.

To bridge the microscopic and macroscopic scales, we introduce an asymptotic framework inspired by \citet{Rosenbaum2018}.\footnote{Our setup is similar to that of \citet{Rosenbaum2018} in the sense that we work with a sequence of processes that become increasingly active as \(T\) increases. The difference is that roughness arises there from a nearly unstable Hawkes kernel, whereas in our model it arises from the behavior of the trawl function near zero.}
For each \(T\in\mathbb N\), we work on a probability space
$(\Omega^T,\mathcal F^T,\mathbb P^T)$,
on which \(L^T\) is an integer-valued, time-homogeneous L\'evy basis on \([0,T]\times\mathbb R\). 
Let \(L_1^T\) denote the associated L\'evy seed, and let \(\nu^T\) denote its L\'evy measure. 
For each \(T\), the trawl component is generated by a \(T\)-dependent trawl function \(d^T\), with associated trawl sets
$A_t^T = \{(x,s):s\leq t,\ b\leq x<d^T(t-s)\}$ for $t\geq 0$.
We equip the probability space with the completed right-continuous filtration \((\mathcal F_t^T)_{t\geq 0}\) generated by \(\{L^T(A_s^T):0\leq s\leq t\}\), \(\{L^T(B_s):0\leq s\leq t\}\), and \(V_0^T\).

Increasing $T$ corresponds to zooming out, so that tick changes occur at an increasingly rapid rate. To capture this behaviour, we let the activity of the L\'evy basis increase linearly in $T$.

\begin{assumption}\label{assumption:inensity_relation}
For each \(T\in\mathbb N\) and \(y\in\mathbb Z\setminus\{0\}\), the L\'evy measures \(\nu^T\) and \(\nu^1\) satisfy
$\nu^T(y)=T\nu^1(y)$.
\end{assumption}

The trawl functions are also \(T\)-dependent. The next assumption requires \(d^T\) to converge monotonically to a limiting trawl function \(d\), while ensuring that each finite-\(T\) trawl set is bounded and has finite Lebesgue measure.

\begin{assumption}\label{assumption:TrawlFunConv}
For each \(T\in\mathbb N\), let \(d^T:[0,\infty)\to[b,T]\) be a non-increasing function satisfying
\[
    \int_0^\infty \bigl(d^T(s)-b\bigr)\,ds<\infty,
    \qquad
    \lim_{s\to\infty}d^T(s)=b,
    \qquad
    d^T(0)\leq T.
\]
Assume further that there exists a non-increasing function
\(d:(0,\infty)\to[b,\infty)\), with \(d-b\) integrable, such that
$d^T(s)-d^T(t)\leq d(s)-d(t)$ for $0<s\leq t$,
for every \(T\). Finally, for every \(B\in\mathcal B(\mathbb R_+)\),
\[
    \int_B \bigl(d^T(s)-b\bigr)\,ds
    \uparrow
    \int_B \bigl(d(s)-b\bigr)\,ds,
    \qquad T\to\infty.
\]
\end{assumption}

\begin{remark}
Although Assumption~\ref{assumption:TrawlFunConv} may appear technical, it is
not restrictive. All it requires is a valid limiting trawl function, that is,
a non-increasing $d$ with $d-b$ integrable, which is then capped at level $T$
by taking $d^T = d\wedge T$. The cap itself is natural rather than
restrictive, since the L\'evy basis $L^T$ lives on $[0,T]\times\mathbb R$.
Thus, $d^T$ is simply a truncated version of the limiting trawl function, with
the truncation level increasing as $T\to\infty$;
\Cref{prop:example_trawl_fun} verifies the assumption for the trawl function
used in the remainder of the paper.
\end{remark}

For each fixed \(T\in\mathbb N\), the microscopic price process is
\begin{equation}\label{eq:MiscroscopicModel}
    P_t^T
    =
    V_0^T
    +
    L^T(A_t^T)
    +
    L^T(B_t),
    \qquad t\geq 0.
\end{equation}
We are going to study the asymptotic behavior of \(P^T=(P_t^T)_{t\geq 0}\) as \(T\to\infty\).

\begin{figure}[t]
    \centering
    \caption{Illustration of $T\to\infty$ in the asymptotic framework.}
    \label{fig:limit_illustration}

    \begin{subfigure}[t]{0.32\textwidth}
        \centering
        \caption*{Panel A: $T=1$}
        \includegraphics[width=\linewidth]{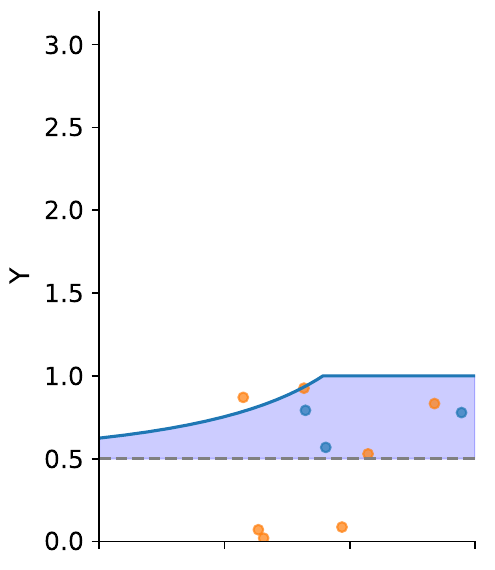}
    \end{subfigure}\hfill
    \begin{subfigure}[t]{0.32\textwidth}
        \centering
        \caption*{Panel B: $T=2$}
        \includegraphics[width=\linewidth]{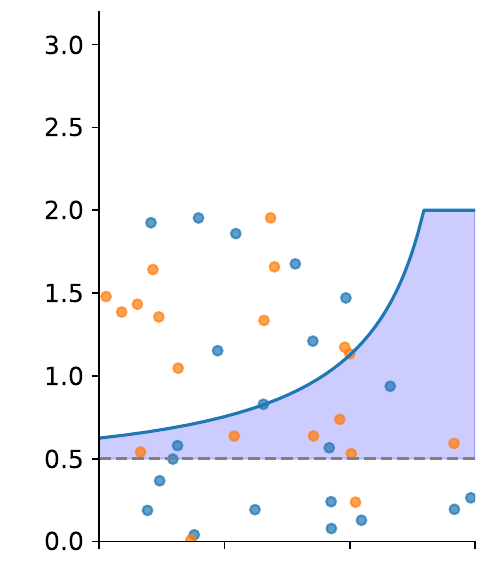}
    \end{subfigure}\hfill
    \begin{subfigure}[t]{0.32\textwidth}
        \centering
        \caption*{Panel C: $T=3$}
        \includegraphics[width=\linewidth]{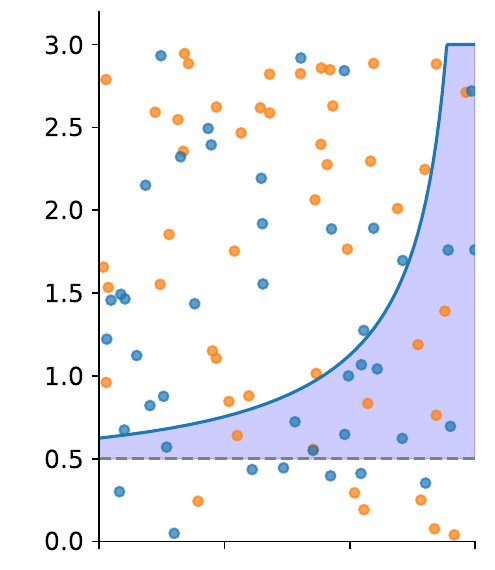}
    \end{subfigure}

    \begin{scriptsize}
    \parbox{\textwidth}{\emph{Note.} In this illustration, we are using a Skellam basis and setting $b=0.5$. The trawl function is given by \eqref{eq:ExampleTrawlFun} with $d^T=d\wedge T$. Blue ticks are upticks while orange dots are downticks, where each tick is $\pm1$.}
    \end{scriptsize}
\end{figure}

\begin{remark}
The asymptotic regime has two effects. First, by Assumption~\ref{assumption:inensity_relation}, the activity of the L\'evy basis increases linearly with \(T\). Thus, as \(T\to\infty\), tick arrivals occur at an increasing rate, reflecting the interpretation that the macroscopic scale aggregates many microscopic price changes. Second, the trawl functions \(d^T\) approach a limiting trawl function \(d\), while the cap \(d^T(0)\leq T\) is allowed to diverge. In the rough specifications considered below, this produces a trawl set that becomes increasingly steep near zero. Consequently, many tick changes enter the trawl component and leave it almost immediately, creating the rapid reversals that give rise to rough noise in the macroscopic limit. These two effects are illustrated in \Cref{fig:limit_illustration}.
\end{remark}

To derive the convergence of the price process, we will require a number of assumptions on the moments of the Lévy seed. As argued in \cite{Barndorff2012}, we may represent the Lévy process associated to the Lévy seed in terms of the difference of two, independent, discrete subordinators, $\tilde{L}_t^T$ and $\bar{L}_t^T$, whose Lévy measures $\tilde{\nu}^T$, $\bar{\nu}^T$ are the restrictions of $\nu^T$ to respectively the positive and negative half axes. We will state the assumptions on the Lévy seed in terms of these.
\begin{assumption}\label{assumption:levy_moments}
        We assume that for all $T$, $\bar{L}_1^T$ is not the degenerate random variable $0$, that $\kappa_1\bigl(\tilde{L}_1^T \bigl) = \kappa_1\bigl(\bar{L}_1^T \bigl)$ and $\kappa_2\bigl(\tilde{L}_1^T \bigl) = \kappa_2\bigl(\bar{L}_1^T \bigl)$, and that $\mathbb{E}\bigl(\vert L_1^1 \vert^3\bigr) < \infty$.
\end{assumption}
\begin{remark}
Assumption \ref{assumption:levy_moments} is not very strict. It simply enforces that we expect there to be no drift in the increments, which can be seen as a \emph{no arbitrage} assumption \citep{Rosenbaum2018}, that the volatility of up- and down ticks should be the same, and that the seed has a finite third absolute moment, which we use in the proof of \Cref{thm:ConvergenceModel}.
\end{remark}

Finally, we will make the very mild assumptions that our efficient price process initially is almost surely finite and does not depend on the frequency at which we are looking at the data. 
\begin{assumption}\label{assumption:v0}
     The distribution of $V_0^T$ does not depend on $T$, and it is almost surely finite.     
\end{assumption}

\subsection{Macroscopic Limit of the Price Model}\label{subsec:macro_limit}

In this subsection, we state the main convergence result for the microscopic price process. The result shows that, after rescaling, the permanent component converges to Brownian motion, while the trawl component converges to a Gaussian moving average. The proof relies on auxiliary convergence results for trawl processes, which are stated in online Appendix~B.

We first impose a condition linking the limiting trawl function to the kernel
of the Gaussian moving-average process that appears in the limit. These
processes take the form $Y_t=\int_{-\infty}^t g(t-s)\,\mathrm d W_s$ for
$t\in\mathbb R$, where \(W\) is a Brownian motion on $\mathbb{R}$ and \(g\) is
a square-integrable kernel.

To describe the local regularity of such processes, we recall the notion of
roughness used throughout the paper. Let \(U=(U_t)_{t\geq 0}\) be a continuous
process with stationary increments. Following \citet{Bennedsen2021}, we say
that \(U\) has roughness index \(\alpha\) if its second-order variogram
satisfies $\mathbb E[(U_h-U_0)^2]=h^{2\alpha+1}L(h)$ for $h>0$, where $L$ is
continuously differentiable, slowly varying at zero,\footnote{By a function
that is slowly varying at zero, we mean a function with the property that for
any $a>0$, $\lim_{h\to0}\frac{L(ah)}{L(h)}=1$.} and bounded away from zero in
a neighbourhood of zero. The index $\alpha$ describes the local regularity of
the sample paths, with smaller values of $\alpha$ corresponding to rougher
local behaviour. Brownian motion has roughness index $0$, and processes with
$\alpha<0$ are rougher than Brownian motion. For a Gaussian moving average,
the roughness is determined by the local behavior of the kernel near zero. In
particular, if $g(x)$ behaves like $x^{\alpha}$ as $x\downarrow 0$, then $Y$
has roughness index $\alpha$ \citep{BLP2021}. The following assumption
specifies the relation between the limiting trawl function \(d\) and the
kernel \(g\).

\begin{assumption}\label{assumption:trawl_fun2}
     The trawl function $d$ is given by $d(h) =b  - \frac{d}{dh} \int_0^\infty g(s) g(h+s) ds$ for every $h>0$, for some $g \in L^2\left({\mathbb{R}}\right)$.
\end{assumption}
\begin{remark}\label{rem:kernelAndtrawl}
The representation in \Cref{assumption:trawl_fun2} does not by itself ensure that $d$ is monotone or that $d(h)\geq b$. Writing $R(h):=\int_0^\infty g(s)g(s+h)ds$,
we have $d(h)=b-R'(h)$. Hence $d$ is non-increasing whenever (R) is convex, and $d(h)\geq b$ whenever $R'(h)\leq 0$. These properties hold, for example, if $g\in C^2((0,\infty))$ is nonnegative, non-increasing, and convex, with sufficient tail integrability to justify differentiating under the integral sign. In derivative form, this sufficient condition may be expressed as
$g\geq 0, g'\leq 0, g''\geq 0$
on $(0,\infty)$, together with the relevant integrability conditions away from the origin.

\end{remark}

For the functional convergence statement, we also record the following sufficient condition.

\begin{assumption}\label{assumption:trawl_fun} 
We assume that the trawl function, $d$, is monotone, and that given any $r,s,t \in \mathbb{R}$ such that $r \le s \le t$ we have $Leb \left( A_t \backslash A_s\right) \le C \left( t-s \right)^{1/2 + \varepsilon /2}$ and $Leb \bigl( \tilde{B}_{t,s,r} \bigl) \le C^\prime \left(t-r\right)^{1 + \varepsilon}$, where $C,C^\prime \in \left(0, \infty \right)$, $\varepsilon >0$ and $\tilde{B}_{t,s,r} = A_s \backslash A_t \backslash A_r$.    
\end{assumption}

\begin{remark}
\Cref{assumption:trawl_fun} is used only for tightness; the finite-dimensional convergence does not rely on it. Its role is to provide deterministic geometric increment bounds tailored to Theorem $13.5$ of \citet{Billingsley1999}. The first bound controls single changes of the moving trawl set. The second controls the middle-only region
$\tilde B_{t,s,r}:=A_s\setminus(A_t\cup A_r)$,
which is present at time $s$ but absent at both neighbouring times $r$ and $t$. For a trawl process of the form $X_u=L(A_u)$, this region enters the adjacent increments $X_s-X_r$ and $X_t-X_s$ with opposite signs, and can therefore create short-time oscillations. Controlling its Lebesgue measure is precisely what allows us to obtain the adjacent-increment probability bounds required for tightness.
\end{remark}

We can now state the main limit theorem for the full price process.

\begin{theorem}\label{thm:ConvergenceModel}
Under Assumptions \ref{assumption:inensity_relation}-\ref{assumption:trawl_fun2}, as $T\rightarrow \infty $, the rescaled price process
\begin{equation*}
    \frac{1}{\sqrt{T\kappa_2\left(L_1^1 \right)}}  P^T_{\centerdot}=  \frac{1}{\sqrt{T\kappa_2\left(L_1^1 \right)}}  \left(V_{0}^T+L^T\left(A_{\centerdot}^T\right)+L^T\left(B_{\centerdot}\right) \right),
\end{equation*}
converges in finite dimensional distribution towards
\begin{equation*}
    P_\centerdot = \sqrt{b}W_\centerdot + \int_{-\infty}^\centerdot g(t-s)d\tilde{W}_s,
\end{equation*}
where $W$ and $\tilde{W}$ are two independent, standard Brownian motions, and the kernel function $g$ is the function introduced in Assumption \ref{assumption:trawl_fun2}. 

If, in addition, Assumption \ref{assumption:trawl_fun} holds and the limiting trawl function is bounded by a finite constant $K$, then the convergence holds in the Skorokhod topology in $[0,\infty)$.\footnote{In this paper, the Skorokhod topology always refers to the $J_1$ topology, which is covered in detail in \cite{Jacod2013}} 
\end{theorem}

\begin{remark}
    Since the limiting process has continuous sample paths, the functional
    convergence statement is equivalent to functional convergence in the locally
    uniform topology, see Proposition~VI.1.17 of \citet{Jacod2013}.
\end{remark}

The first part of \Cref{thm:ConvergenceModel} is the main convergence result used in the paper. It shows that the permanent component converges to Brownian motion, while the fleeting trawl component converges to a Gaussian moving average. The second part gives a sufficient condition under which the convergence can be strengthened from finite-dimensional convergence to functional convergence. As we will see in the following subsection, this sufficient condition is satisfied in the non-rough case but fails for the rough trawl functions considered below. Thus, in the rough-noise case, the link between the microscopic and macroscopic models is established through finite-dimensional convergence. This is the same type of convergence that is obtained in the rough volatility case in \cite{Rosenbaum2018}.

\begin{subsection}{A Rough Noise Model}\label{Rough Noise Model}
We now construct a trawl function whose macroscopic limit attains any desired roughness level. The target is the rough-noise kernel \eqref{eq:chong_kernel} of \citet{ChongEtAl2021}. Since $t\mapsto t^{H-1/2}$ is not square-integrable for $H\in(0,1/2)$, this kernel violates \Cref{assumption:trawl_fun2} and cannot serve as our limit. Moreover, with their volatility process $\rho$ held constant, their construction yields a truncated, and hence non-stationary, Gaussian moving average, whereas \Cref{thm:ConvergenceModel} yields a non-truncated stationary limit.

We can nonetheless match the local behavior near zero, and therefore the roughness of the sample paths. Specifically, we consider a Gaussian moving-average process with the gamma kernel $g(x)=x^\alpha\exp(-\lambda x),\quad x>0,\;\alpha\in(-1/2,1/2),\;\lambda>0$
where $\alpha$ is the roughness index. For this kernel function, we find
\begin{proposition}\label{prop:example_trawl_fun_def}
    For $g$ given by the gamma kernel, the function
    \begin{equation}\label{eq:ExampleTrawlFun}
        d(h) = b + \frac{\Gamma(\alpha+1)}{\sqrt{\pi}}\,\lambda\left(\frac{h}{2\lambda}\right)^{\alpha+1/2}K_{\alpha-1/2}(\lambda h)
    \end{equation}
    satisfies \Cref{assumption:trawl_fun2}.
\end{proposition}
The resulting limiting process can thus exhibit any roughness level between $-1/2$ and $1/2$, and is essentially the kernel function of \cite{ChongEtAl2021}, but with an exponential dampening function.

The next proposition establishes that $d$ in \eqref{eq:ExampleTrawlFun} satisfies the regularity required by \Cref{thm:ConvergenceModel} and characterizes its local behavior near zero, which drives the macroscopic roughness.

\begin{proposition}\label{prop:example_trawl_fun}
    For the trawl function $d$ of \Cref{prop:example_trawl_fun_def}, the following hold:
    \begin{enumerate}
        \item For  $\alpha\in(-1/2,0]$,  $d^T=\min(d,T)$ satisfies \Cref{assumption:TrawlFunConv}.
        \item For  $\alpha\in(-1/2,0)$, $d(h)\to\infty$ as $h\to 0$. For $\alpha=0$, $d(h)=b+\tfrac{1}{2}e^{-\lambda h}$. In the latter case, if $b\le\tfrac{1}{2}$ then $d^1=\min(d,1)=d$.
    \end{enumerate}

\end{proposition}

The proposition confines the relevant parameter range to $\alpha\in(-1/2,0]$, with macroscopic noise that is Brownian at $\alpha=0$ and rough for $\alpha\in(-1/2,0)$.

The dichotomy reflects how tick changes enter and exit the trawl set. At $\alpha=0$ the trawl function is bounded near zero, and the rescaling produces a Brownian limit that mirrors the efficient-price component. For $\alpha<0$ the asymptote at zero allows ticks to fall in and out of the trawl set almost instantaneously, generating the rough limit. This is also why our asymptotic framework requires a trawl function that becomes unbounded as $T\to\infty$, rather than the bounded-trawl setup of \citet{PPSV2021}: a bounded trawl function cannot yield a rough process in the limit. Part 1 of \Cref{prop:example_trawl_fun} is also used in the proofs of the estimator's properties in \Cref{sec:estimation}.

Finally, the mode of convergence in \Cref{thm:ConvergenceModel} depends on $\alpha$. Functional convergence requires \Cref{assumption:trawl_fun}, which holds only at $\alpha=0$.

\begin{proposition}\label{prop:LebMeas}
    For the trawl function in \eqref{eq:ExampleTrawlFun}, \Cref{assumption:trawl_fun} is satisfied at $\alpha=0$ and violated for every $\alpha\in(-1/2,0)$.
\end{proposition}
\begin{remark}
It can in fact be shown that any monotone trawl functions $d$ satisfying \Cref{assumption:trawl_fun2} and giving rise to roughness in the limit, violates the bound in \Cref{assumption:trawl_fun}. To see this, note that the limiting Gaussian moving average has autocovariance $R(h) = \operatorname{Cov}(Y_t,Y_{t+h}) = \int_0^\infty g(s)g(s+h)ds,$
so its variogram is $\mathbb{E}[(Y_h-Y_0)^2] = 2\bigl(R(0)-R(h)\bigr)$.
By our definition of roughness, $Y$ is rough when
$ 2\bigl(R(0)-R(h)\bigr) \sim C h^{2\alpha+1}L(h)$, as $h\downarrow0$,
with $\alpha<0$. This irregular behaviour implies $d(h)-b\sim C' h^{2\alpha}L(h)$ as, $ h\downarrow0$. Since $\alpha<0$, this explodes at the origin. This singularity is precisely what causes $\operatorname{Leb}(\tilde B_{t,s,r})$ to decay too slowly to satisfy the bound in \Cref{assumption:trawl_fun}.

\end{remark}

\begin{remark}\label{rem:sv_extension}
The convergence in \Cref{thm:ConvergenceModel} extends to a stochastic-volatility version, obtained by time-changing $L^T(B_t)$ in \eqref{eq:MiscroscopicModel} with an independent subordinator $\tau(t)=\int_0^t\upsilon_u^2\,du$: the same rescaling yields a limiting price process of the form $\sqrt{b}\,W_{\tau(\cdot)}+\int_{-\infty}^{\cdot}g(t-s)\,d\tilde W_s$, a Bachelier price with stochastic volatility plus the same rough noise. We work with the homoskedastic version below, while \citet{ChristensenEtAL2023} develop estimation for the SV setting.
\end{remark}


\end{subsection}

\section{Estimation and Inference}\label{sec:estimation}

From this section onward, we work with the $T=1$ price process, i.e.\ the one generated by a Lévy basis on $[0,1]\times\mathbb{R}$. We refer to this as the standard-time scale, representing the highest observation frequency in our data. Since $T$ is fixed henceforth, we suppress the $T$-superscripts to ease notation.

We derive a generalized method of moments (GMM) estimator for the parameters of the model in \Cref{Rough Noise Model}, characterize its asymptotic behavior, and develop a formal test for the presence of rough noise. All proofs are deferred to online Appendices~A and~C. The parameters are $\theta = (\alpha,\lambda,b,\kappa_2(L_1))^\top$, where $\alpha$ and $\lambda$ parameterize the trawl function $d$ and $\kappa_2(L_1)$ is the variance of the Lévy seed. The roughness index $\alpha$ is the central object of interest, since it governs whether the noise is rough.

Our estimator is semiparametric. The model accommodates a wide range of Lévy seeds, but our interest is in the roughness of the noise, which is controlled by the trawl function rather than by the Lévy seed itself. We therefore neither assume a parametric form for the Lévy seed nor estimate its parameters beyond the second cumulant $\kappa_2(L_1)$. The first cumulant, corresponding to the mean, is set to zero by \Cref{assumption:levy_moments}.

\subsection{Estimation Method}

We propose a GMM estimator similar to that of \citet{Shephard2017}. For $h>0$, let $\Delta P_t^h := P_{t+h} - P_t$ denote the price increment over a time step of length $h$ at time $t$. For each $m \in \mathbb{N}$, let $D_m \subset \mathbb{N}$ be a finite set of $m$ positive integers, written $D_m = \{l_1,\dots,l_m\}$, which index the lags used in estimation. Define the vector of increments $Y_t^{(m)} = \left( \Delta P_{t\delta}^{l_1 \delta},\Delta P_{t\delta}^{l_2 \delta},\dots,\Delta P_{t\delta}^{l_m \delta}\right)$ for $t = 1,\dots,n-l_m$, where $\delta = 1/n$ for some $n \in \mathbb{N}$. Let $\Theta$ denote the parameter space and set $D(k,\theta) := \mathbb{E}\!\left[\left(\Delta P_0^{l_k\delta}\right)^2\right]$, $k = 1,\dots,m$. The sample moment function is $g_{n, m}(\theta) = \frac{1}{n-l_m} \sum_{t=1}^{n-l_m} h\!\left(Y_t^{(m)}, \theta\right)$, where $h:\mathbb{R}^m \times \Theta \to \mathbb{R}^m$ stacks the centered second moments, $h\!\left(Y_t^{(m)}, \theta\right) = \left((\Delta P_{t\delta}^{l_1 \delta})^2 - D(1, \theta),\dots,(\Delta P_{t\delta}^{l_m \delta})^2 - D(m, \theta)\right)^{\top}$. The GMM estimator of $\theta_0$ is 
\begin{equation*}
    \widehat{\theta}_{0,\mathrm{GMM}}^{n, m} = \operatorname*{argmin}_{\theta \in \Theta}\, g_{n, m}(\theta)^{\top} A_{n, m}\, g_{n, m}(\theta),
\end{equation*}
where $A_{n, m}$ is the diagonal weighting matrix with $i$th diagonal entry $1/(l_i\delta)^2$. The weights $1/(l_i\delta)^2$ are (approximate) inverse-variance weights under the null hypothesis $\alpha = 0$, where $\mathbb{E}[(\Delta P_t^h)^2]$ scales linearly in $h$. This is not efficient in general, but consistency of $\widehat\theta_{0,\mathrm{GMM}}^{n,m}$ does not depend on $A_{n,m}$. Consistency and the asymptotic distribution of $\widehat\theta_{0,\mathrm{GMM}}^{n,m}$ are established in the next subsection.

\subsection{Asymptotic Properties}\label{PropertiesEstimator}
Since we work with integer-valued Lévy bases, the corresponding Lévy seed has characteristic triplet $(\gamma, 0, \eta)$, where $\gamma = \int_{\mathbb{R}} \xi \mathbb{I}_{[-1,1]}(\xi)\,\eta(d\xi) = \sum_{\xi=-1}^{1} \xi\,\eta(\xi)$. We impose the following assumptions.

\begin{assumption}\label{assumption:AssumptionsConsistency}
    \begin{enumerate}
        \item $\int_{|\xi|>1} |\xi|^{4+\delta}\,\eta(d\xi) < \infty$ for some $\delta > 0$. \label{assumption:consistency1}
        \item $\Theta$ is compact and contains $\theta_0$. \label{assumption:consistency2}
    \end{enumerate}
\end{assumption}

\begin{assumption}\label{assumption:AsymptoticIdentification}
    For each $m \in \mathbb{N}$, $D_m \subseteq D_{m+1}$, and $l_m \to \infty$ as $m \to \infty$.
\end{assumption}

\begin{remark}
    Part 1 of \Cref{assumption:AssumptionsConsistency} is a mild moment condition. Part 2 is similarly mild and standard in the GMM consistency and CLT literature; parameter bounds are typically imposed during optimization. \Cref{assumption:AsymptoticIdentification} ensures that the lag indices in $D_m$ grow without bound, which is needed for the identification result below.
\end{remark}

We first establish asymptotic identification of the parameter vector.

\begin{lemma}\label{lemma:AsymIdentification}
    Under Assumptions \ref{assumption:levy_moments}, \ref{assumption:AssumptionsConsistency}, and \ref{assumption:AsymptoticIdentification}, the parameter vector is asymptotically identifiable: if $h_t^{(m)}(\theta) := \mathbb{E}\!\left[h\!\left(Y_t^{(m)}, \theta\right)\right]$, then $h_t^{(m)}(\theta) = 0$ for all $t$ and $m \in \mathbb{N}$ if and only if $\theta = \theta_0$.
\end{lemma}

Asymptotic identification, in the sense of Definition 3.3 of \citet{Stelzer2015} and \citet{Curato2019}, ensures that the parameters are nearly identified for sufficiently large $m$. Since exact identification is needed for consistency, we strengthen \Cref{lemma:AsymIdentification} to an assumption; the simulation study in \Cref{sec:sim} provides further evidence that the parameters are identified in practice.

\begin{assumption}\label{assumption:Identification}
    For all sufficiently large $m$, $h_t^{(m)}(\theta) = 0$ for all $t$ if and only if $\theta = \theta_0$.
\end{assumption}

In the remainder of this subsection we assume that $m$ is sufficiently large.

\begin{theorem}\label{thm:Consistency}
    Under Assumptions \ref{assumption:levy_moments}, \ref{assumption:AssumptionsConsistency}, and \ref{assumption:Identification}, the GMM estimator is weakly consistent: $\widehat{\theta}_{0, \mathrm{GMM}}^{n, m} \overset{\mathbb{P}}{\to} \theta_0$ as $n \to \infty$.
\end{theorem}

We next derive the asymptotic distribution, distinguishing the case where $\theta_0$ lies in the interior of $\Theta$ from the boundary case $\alpha_0=0$ that is relevant for testing non-rough noise. In the interior a standard $\sqrt{n}$ central limit theorem applies; on the boundary we invoke \citet{Andrews2002}, so the rate remains $\sqrt{n}$ but the limit is a one-sided normal with a point mass at zero.

\begin{theorem}\label{thm:CLT}\label{thm:CLT_a0}
    Suppose Assumptions \ref{assumption:levy_moments}, \ref{assumption:AssumptionsConsistency}, and \ref{assumption:Identification} hold, and set
    \begin{equation*}
        \Sigma_a = \sum_{l \in \mathbb{Z}} \operatorname{Cov}\!\left(h\!\left(Y_0^{(m)}, \theta_0\right),\, h\!\left(Y_l^{(m)}, \theta_0\right)\right),
        \qquad
        M = \left(G_0^{\top} A G_0\right)^{-1} G_0^{\top} A,
    \end{equation*}
    with $G_0 = \left.\dfrac{\partial h\!\left(Y_t^{(m)}, \theta\right)}{\partial \theta^{\top}}\right|_{\theta=\theta_0}$ and $A = A_{n,m}$. Then, as $n \to \infty$:
    \begin{enumerate}
        \item If $\theta_0 \in \interior{\Theta}$, then
            \[
                \sqrt{n}\!\left(\widehat{\theta}_{0, \mathrm{GMM}}^{n, m} - \theta_0\right) \overset{d}{\to} \mathrm{N}\!\left(0,\, M \Sigma_a M^{\top}\right).
            \]
        \item If $\alpha_0 = 0$ and $b_0 \in (0,1)$, then
            \[
                \sqrt{n}\,\widehat{\alpha} \overset{d}{\to} Z \mathbbm{1}_{\{Z \le 0\}}, \qquad Z \sim \mathrm{N}\!\left(0,\, H M \Sigma_a M^{\top} H^{\top}\right),
            \]
            where $H = (1, 0, \ldots, 0)$ and $\widehat{\alpha}$ is the $\alpha$-coordinate of $\widehat{\theta}_{0,\mathrm{GMM}}^{n,m}$.
    \end{enumerate}
\end{theorem}

\begin{remark}\label{rem:feasible_test}
\Cref{thm:CLT_a0} yields a test of $H_0:\alpha=0$ (no rough noise) against $H_1:\alpha<0$, but it is infeasible since the asymptotic variance of $\widehat\alpha$ is unknown. Replacing it with the consistent estimator $\widehat\sigma_\alpha^{n,k}$ constructed in the online supplement gives the feasible statistic $T_n=\sqrt{n/\widehat\sigma_\alpha^{n,k}}\,\widehat\alpha$, which satisfies $T_n\overset{d}{\to} Z\mathbbm{1}_{\{Z\le0\}}$, $Z\sim\mathrm N(0,1)$, under $H_0$. The test rejects at level $\gamma$ when $T_n$ falls below the $\gamma$-quantile of $\mathrm N(0,1)$ (e.g.\ $-1.645$ at $5\%$); the interior-$\alpha_0$ test and all proofs are found in Appendix~C.
\end{remark}

\section{Simulation Study}\label{sec:sim}
To assess the finite-sample performance of the GMM estimator, we conduct a Monte Carlo study, using the trawl function in \eqref{eq:ExampleTrawlFun} so that the simulated process resembles real tick-by-tick data.

We take $\mathcal{T}\in\{5850,11700,23400\}$ time periods, interpreted as seconds. This corresponds to a quarter trading day, half a trading day, and a full trading day, respectively. The grid spacing is $\delta=0.1$, giving one observation every tenth of a second. The underlying Lévy basis is Skellam with intensity $\nu\bl\mathbb{Z}\backslash\{0\}\br=6.28$ per second, so each tick is $\pm 1$, which means we expect just over $6$ price changes per second. This corresponds to our empirical average. 

 The GMM moment conditions use $m=100$ lags $D_m=\{l_1,\dots,l_{100}\}$ ranging from $l_1=1$ to $l_{100}=100$ (in units of $\delta$), equidistantly-spaced. The weighting matrix is the diagonal $A_{n,m}$ from \Cref{sec:estimation}, and the long-run covariance estimator $\widehat{\Sigma}_a^k(\widehat\theta^n_{\mathrm{full}})$ truncates lag covariances at $k=1000$. We simulate $R=1000$ replications and report the results in \Cref{tab:estimates_meanDGP}.



\begin{table}[H]
\centering
\caption{Parameter estimates with empirically-anchored ``mean'' DGP.}
\label{tab:estimates_meanDGP}
\small
\begin{tabular}{lrrrrr}
    \toprule
    Parameter & Value & & \multicolumn{3}{c}{Gamma model} \\ \cmidrule(lr){4-6}
    & & & \multicolumn{1}{c}{$\mathcal{T}=5850$} & \multicolumn{1}{c}{$\mathcal{T}=11700$} & \multicolumn{1}{c}{$\mathcal{T}=23400$} \\
    \midrule
    \multicolumn{6}{l}{Panel A} \\
    $\alpha$   & $-0.400$ & & $-0.368$ $(0.141)$ & $-0.382$ $(0.115)$ & $-0.395$ $(0.094)$ \\
    $b$        & $0.740$  & & $0.739$ $(0.023)$ & $0.739$ $(0.016)$ & $0.740$ $(0.012)$ \\
    $\lambda$  & $4.110$  & & $4.064$ $(1.222)$ & $4.078$ $(0.978)$ & $4.141$ $(0.798)$ \\
    $\kappa_2$ & $6.280$  & & $6.288$ $(0.128)$ & $6.284$ $(0.088)$ & $6.289$ $(0.064)$ \\
    \addlinespace
    \multicolumn{6}{l}{Panel B} \\
    $\alpha$   & $-0.300$ & & $-0.309$ $(0.155)$ & $-0.313$ $(0.130)$ & $-0.311$ $(0.105)$ \\
    $b$        & $0.740$  & & $0.738$ $(0.023)$ & $0.739$ $(0.019)$ & $0.739$ $(0.011)$ \\
    $\lambda$  & $4.110$  & & $4.385$ $(1.586)$ & $4.348$ $(1.291)$ & $4.258$ $(1.018)$ \\
    $\kappa_2$ & $6.280$  & & $6.291$ $(0.125)$ & $6.289$ $(0.089)$ & $6.288$ $(0.062)$ \\
    \addlinespace
    \multicolumn{6}{l}{Panel C} \\
    $\alpha$   & $-0.177$ & & $-0.208$ $(0.150)$ & $-0.197$ $(0.114)$ & $-0.188$ $(0.083)$ \\
    $b$        & $0.740$  & & $0.739$ $(0.023)$ & $0.739$ $(0.016)$ & $0.740$ $(0.012)$ \\
    $\lambda$  & $4.110$  & & $4.608$ $(1.959)$ & $4.406$ $(1.427)$ & $4.266$ $(0.998)$ \\
    $\kappa_2$ & $6.280$  & & $6.285$ $(0.120)$ & $6.287$ $(0.082)$ & $6.289$ $(0.063)$ \\
    \addlinespace
    \multicolumn{6}{l}{Panel D} \\
    $\alpha$   & $-0.080$ & & $-0.124$ $(0.128)$ & $-0.099$ $(0.083)$ & $-0.088$ $(0.056)$ \\
    $b$        & $0.740$  & & $0.739$ $(0.025)$ & $0.740$ $(0.016)$ & $0.739$ $(0.015)$ \\
    $\lambda$  & $4.110$  & & $4.816$ $(2.216)$ & $4.406$ $(1.330)$ & $4.230$ $(0.855)$ \\
    $\kappa_2$ & $6.280$  & & $6.287$ $(0.121)$ & $6.286$ $(0.080)$ & $6.289$ $(0.064)$ \\
    \addlinespace
    \multicolumn{6}{l}{Panel E} \\
    $\alpha$   & $-0.036$ & & $-0.068$ $(0.088)$ & $-0.051$ $(0.053)$ & $-0.044$ $(0.040)$ \\
    $b$        & $0.740$  & & $0.740$ $(0.023)$ & $0.740$ $(0.015)$ & $0.739$ $(0.016)$ \\
    $\lambda$  & $4.110$  & & $4.657$ $(1.918)$ & $4.343$ $(1.047)$ & $4.238$ $(0.782)$ \\
    $\kappa_2$ & $6.280$  & & $6.284$ $(0.116)$ & $6.288$ $(0.078)$ & $6.291$ $(0.070)$ \\
    \addlinespace
    \multicolumn{6}{l}{Panel F} \\
    $\alpha$   & $0.000$  & & $-0.021$ $(0.055)$ & $-0.011$ $(0.031)$ & $-0.007$ $(0.020)$ \\
    $b$        & $0.740$  & & $0.741$ $(0.020)$ & $0.740$ $(0.014)$ & $0.740$ $(0.010)$ \\
    $\lambda$  & $4.110$  & & $4.489$ $(1.428)$ & $4.289$ $(0.798)$ & $4.206$ $(0.538)$ \\
    $\kappa_2$ & $6.280$  & & $6.282$ $(0.128)$ & $6.282$ $(0.086)$ & $6.285$ $(0.063)$ \\
    \bottomrule
\end{tabular}

\begin{scriptsize}
\parbox{\textwidth}{\emph{Note.} Results based on $R=1000$ repetitions. Skellam basis with $\nu(\mathbb{Z}\backslash\{0\}) = 6.28$. Fixed parameters $b=0.74$, $\lambda=4.11$, $\kappa_2=6.28$ from empirical means. Mean estimates with Monte Carlo standard deviations in parentheses.}
\end{scriptsize}
\end{table}

We see that the GMM estimator performs well across roughness levels and sample sizes: bias is negligible and the Monte Carlo standard deviations are small in every setting.

\begin{figure}[H]
    \centering
    \caption{Finite-sample power and null distribution of the rough-noise test.}
    \label{fig:power_curve_H0_dist}

    \begin{subfigure}[t]{0.48\textwidth}
        \centering
        \caption*{Panel A: Power curve}
        \includegraphics[width=\linewidth]{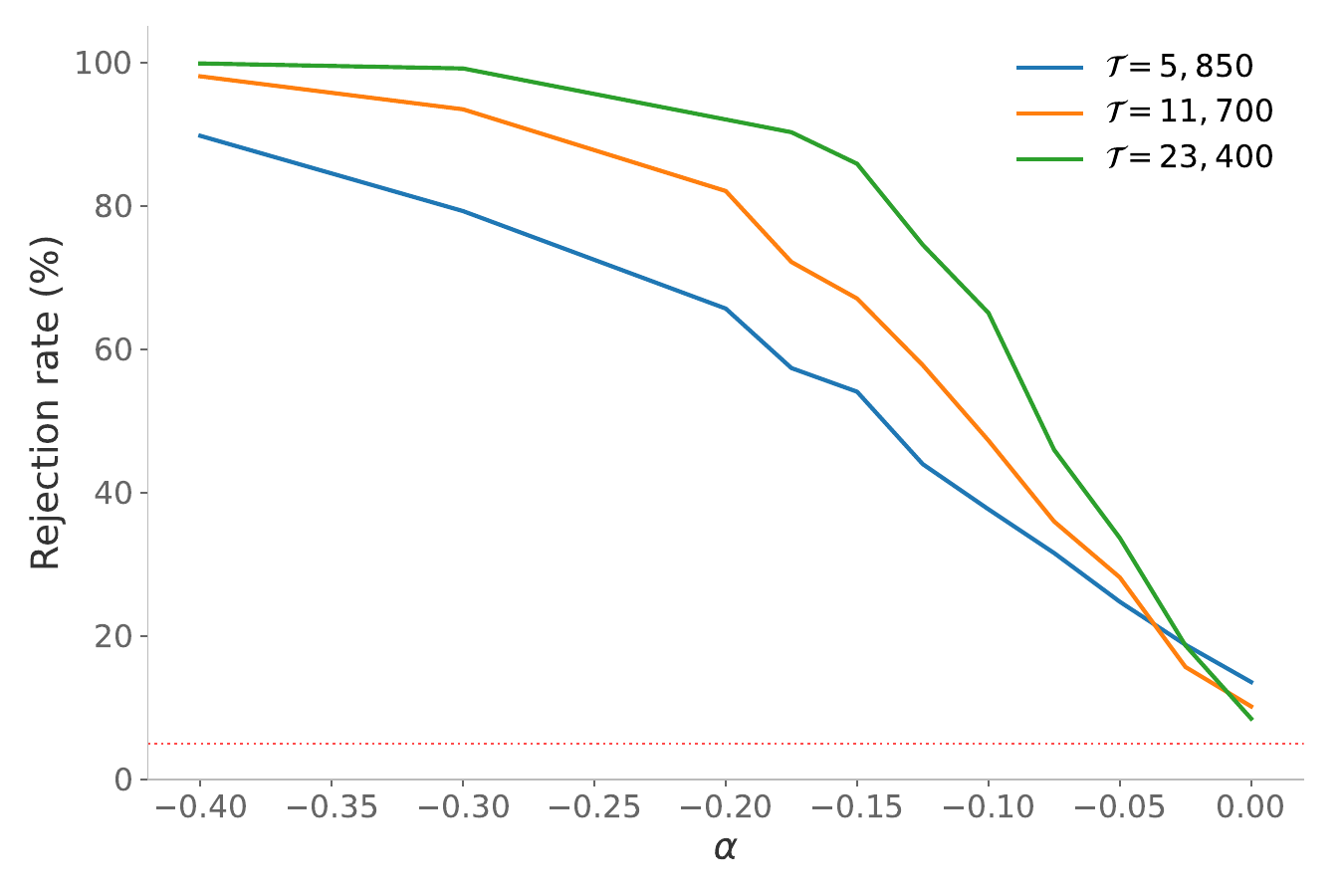}
    \end{subfigure}\hfill
    \begin{subfigure}[t]{0.48\textwidth}
        \centering
        \caption*{Panel B: Empirical CDF under $H_0$}
        \includegraphics[width=\linewidth]{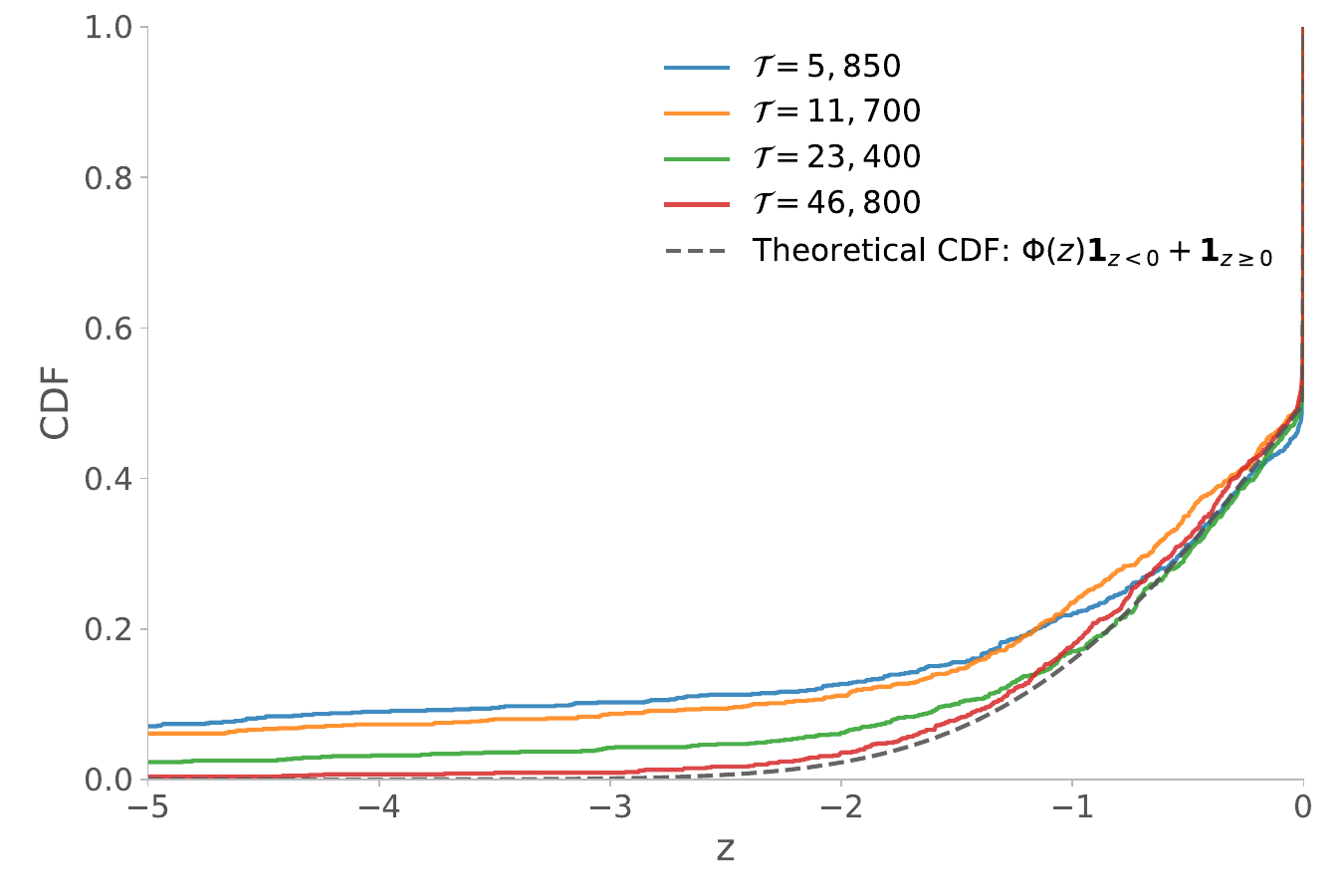}
    \end{subfigure}

    \begin{scriptsize}
    \parbox{\textwidth}{\emph{Note.} Panel A: rejection frequency of the test
    of \Cref{rem:feasible_test} at the $5\%$ nominal level, plotted against the true
    rough-noise index $\alpha$. Each point is based on $R=1000$ Monte Carlo
    replications for $\mathcal{T} \in \{5850,\,11700,\,23400\}$; the point at
    $\alpha=0$ is the empirical size and the remaining points are the
    empirical power under $H_1:\alpha<0$. The horizontal dotted red line
    marks the nominal $5\%$ level and the vertical dotted line marks
    $\alpha=0$. Panel B: empirical CDF of the test statistic under
    $H_0:\alpha_0=0$ based on $R=1000$ replications for
    $\mathcal{T} \in \{5850,\,11700,\,23400,\,46800\}$, compared to the
    theoretical limit CDF $F(x)=\Phi(x)\mathbf{1}_{\{x<0\}}+\mathbf{1}_{\{x\geq 0\}}$
    of the mixed limit $Z\mathbf{1}_{\{Z\leq 0\}}$. The empirical rejection
    frequency approaches the nominal $5\%$ level as $\mathcal{T}$ grows,
    reaching approximately $6\%$ at $\mathcal{T}=46{,}800$.}
    \end{scriptsize}
\end{figure}

Next, we investigate the size and power of the test. We perform $1000$ replications at each of $11$ values of $\alpha$ between $-0.40$ and $0$, testing $H_0:\alpha=0$ at the $5\%$ significance level. The rejection rates are shown in Panel A of \Cref{fig:power_curve_H0_dist}.

The power curve has the expected shape: rejection rates approach $1$ quickly as $\alpha$ moves away from the null. The rejection rate at $\alpha_0=0$ exceeds the nominal $5\%$ level but improves substantially as the sample size grows, suggesting that the size distortion is a finite-sample effect (our largest $\mathcal{T}=23400$ corresponds to a single trading day).

We next examine the distribution of the test statistic under the null $\alpha_0=0$. By \Cref{rem:feasible_test}, $T_n\overset{d}{\to} Z\mathbbm{1}_{\{Z\le 0\}}$ with $Z\sim\mathrm{N}(0,1)$, which places mass $1/2$ at zero and is half-normal on the negative half-line. Panel B of \Cref{fig:power_curve_H0_dist} plots the empirical CDF of $T_n$ for $\mathcal{T}\in\{5850,11700,23400,46800\}$ against the limit CDF $F(x)=\Phi(x)\mathbbm{1}_{\{x<0\}}+\mathbbm{1}_{\{x\ge 0\}}$.

For smaller samples, the empirical distribution is somewhat heavier-tailed than the limit, consistent with the inflated size in Panel A. However, the fit improves substantially as $\mathcal{T}$ grows, suggesting once again that this is a finite-sample phenomenon. We also note that the median of the test statistics, which according to theory should be exactly zero, is zero up to the 4th decimal point for the two larger sample sizes. Overall, the simulations seem to support our theoretical results.

\section{Empirical application}
\label{sec:empirical}

We apply the estimation method from \Cref{sec:estimation} to tick-by-tick transaction data for the constituents of the Dow Jones Industrial Average (DJIA) during 2024, producing a roughness estimate for each ticker-day. The DJIA composition is taken as of the beginning of the year. Motivated by the cross-venue evidence in \Cref{sec:TickModelChoice}, we use trades from all available exchanges. The raw transaction data is otherwise processed following \citet{Barndorff2009}.\footnote{We follow the trade-based cleaning rules of \citet{Barndorff2009}, with two deviations. First, we keep trades from all available exchanges, dropping their single-exchange restriction (rule P3), since the cross-venue evidence in \Cref{sec:TickModelChoice} motivates the all-exchange construction. Second, we omit the quote-based filters, since we do not have access to quote data.}

As discussed in \Cref{sec:TickModelChoice}, in a tick-by-tick model it is natural to interpret microstructure noise as a source of short-run reversals in observed transaction prices.

This motivates two empirical questions. First, how prevalent is rough noise among DJIA constituents in 2024? Second, are rough-noise detections associated with observable trading characteristics that reflect the high-frequency frictions underlying microstructure noise? We address the first by reporting the share of ticker-days for which the test of \Cref{rem:feasible_test} rejects the null of non-rough noise, together with descriptive statistics of the parameter estimates. We address the second by following \citet{Ait2009} in regressing the rough-noise indicator on standard liquidity, activity, and price-level measures.

We use the same estimation configuration as in the simulation study of \Cref{sec:sim}, namely prices sampled at $\delta=0.1$ seconds, $m=100$ equidistantly-spaced lags, the diagonal weighting matrix $A_{n,m}$, and the long-run covariance estimator $\widehat\Sigma_a^k$ truncated at $k=1000$. Since the model parameters are not identifiable in the absence of microstructure noise, we restrict the sample using the HAC Wald pre-test from \Cref{sec:TickModelChoice} for the null of non-negative first-order autocorrelation,
$H_0:\rho(1)\geq 0$,
at the $5\%$ level, retaining only ticker-days for which the null is rejected. Across the full sample, $64\%$ of ticker-day observations exhibit evidence of microstructure noise. For these noisy ticker-days, we estimate the model parameters and apply the test of \Cref{rem:feasible_test}. We find that $20.4\%$ of noisy ticker-days exhibit rough noise, a non-negligible share.




\Cref{tab:parameter_estimate_distribution} reports the empirical distribution of the parameter estimates across noisy ticker-days. The distribution of $\alpha$ is concentrated near zero, with a median of essentially zero and only the most rough $5\%$ of estimates falling below $-0.18$. Conditioning on the subsample where the null of non-rough noise is rejected, the median estimate is $-0.08$.

At first glance, our findings seem to contrast sharply with \citet[Figure~5]{ChongEtAl2021}, who report a stable roughness level over the period 2013--2022 corresponding in our notation to $\alpha \approx -0.20$. The apparent disagreement, however, is consistent with the difference in estimation methodology between the two studies. They estimate based on power variation over a 5-day rolling window, which will asymptotically be dominated by the roughest day in the window. A single day with $\alpha < 0$ pulls $\widehat\alpha$ toward its value even when the surrounding four days are non-rough. Their $\alpha \approx -0.20$ is therefore consistent with rough noise occurring episodically rather than systematically, with rolling estimates reflecting the most rough day in each window rather than a typical day. Our day-by-day estimator unbundles this signal and reveals substantial day-to-day variation in the underlying roughness, with most days exhibiting no roughness at all.\footnote{The estimator occasionally hits the boundary at $\alpha=-1/2$ or $\lambda=0$. These cases coincide with days where $b$ is estimated close to its upper boundary of $1$, on which the trawl component shrinks toward zero and the noise parameters $\alpha$ and $\lambda$ become hard to identify.}

\begin{table}[!htbp]
\centering
\caption{Distribution of parameter estimates.}
\label{tab:parameter_estimate_distribution}
\begin{threeparttable}
\small
\begin{tabular}{lrrrrrrrrr}
\toprule
Parameter & $N$ & Mean & Min & $q_{0.05}$ & $q_{0.25}$ & Median & $q_{0.75}$ & $q_{0.95}$ & Max \\
\midrule
$b$        & 4,765 &  0.742 &  0.007 &  0.441 &  0.678 & 0.775 & 0.845 &  0.916 &   0.969 \\
$\lambda$  & 4,765 &  4.113 &  0.000 &  0.031 &  0.637 & 1.569 & 4.535 & 19.990 &  29.088 \\
$\kappa_2$ & 4,765 &  6.290 &  0.062 &  0.117 &  0.592 & 2.583 & 7.011 & 23.130 & 115.821 \\
$\alpha$   & 4,765 & -0.036 & -0.500 & -0.177 & -0.021 & 0.000 & 0.000 &  0.000 &   0.000 \\
\bottomrule
\end{tabular}
\begin{tablenotes}[flushleft]
\footnotesize
\item \textit{Notes:} The table reports the empirical distribution of parameter estimates across ticker-day estimations.
\end{tablenotes}
\end{threeparttable}
\end{table}

We next study which empirical characteristics explain the presence of rough noise.  \citet{Ait2009} show that the magnitude of market microstructure noise is closely related to stock liquidity, with estimated noise and noise-to-signal ratios lower for more liquid stocks. We follow the spirit of their exercise but replace the estimated noise magnitude with a binary indicator for rough noise.

Specifically, we estimate linear probability models
$ROUGH_{i,d} = \beta_0 + \beta' X_{i,d} + \varepsilon_{i,d}$,
where $ROUGH_{i,d} = \mathbf{1}\{T_{i,d} \le -1.645\}$ is the rough-noise indicator from the test of \Cref{rem:feasible_test}. The critical value corresponds to a 5\% test under the distribution of our test-statistic under the null.



The vector $X_{i,d}$ is chosen to mirror the liquidity measures in \citet{Ait2009} as closely as possible given our data:
\[
\begin{aligned}
X_{i,d}
=
\big[&
\sigma_{i,d},
\text{RollSpread}_{i,d},
LOGTRADESIZE_{i,d},
LOGNTRADE_{i,d}, \\
&
MONTHVOL_{i,d},
LOGP_{i,d},
ILLIQ_{i,d}
\big]^{\top}.
\end{aligned}
\]
Here, $\sigma_{i,d}$ is the intraday realized volatility based on $5$-minute returns, $LOGTRADESIZE_{i,d}$ and $LOGNTRADE_{i,d}$ are the logarithms of the average trade size and of the number of intraday trades, $MONTHVOL_{i,d}$ is a lagged 21-trading-day realized-volatility measure, $LOGP_{i,d}$ is the logarithm of the daily closing price, and $ILLIQ_{i,d}$ is the daily Amihud illiquidity measure \citep{Amihud2002}, computed as absolute daily return divided by dollar volume.

Since we do not have quoted bid--ask spreads, we follow \citet{Roll1984} and proxy the average spread by the Roll spread,
$\text{RollSpread}_{i,d} = 2\sqrt{\max\{-\gamma_{1,i,d},\,0\}}$,
where $\gamma_{1,i,d}$ is the first-order autocovariance of intraday price changes. The Roll spread is a direct function of the negative autocovariance and therefore captures short-run price reversals. As established in \Cref{sec:simple_model}, this is the mechanism through which microstructure noise enters our framework. Rough noise corresponds to this mechanism in its extreme form, with reversals that are frequent, short-lived, and large relative to the price level.

\begin{table}[!htbp]
\centering
\caption{Regression of roughness detection dummy on liquidity and reversal measures}
\label{tab:rough_dummy_liquidity_reversal}
\scriptsize
\setlength{\tabcolsep}{8pt}
\begin{tabular}{l
                S[table-format=-2.3] S[table-format=-1.2] S[table-format=1.2]
                S[table-format=-1.3] S[table-format=-1.2]
                S[table-format=-2.3] S[table-format=-1.2]}
\toprule
& \multicolumn{3}{c}{(1) Individual}
& \multicolumn{2}{c}{(2) Joint excl. Roll}
& \multicolumn{2}{c}{(3) Joint incl. Roll} \\
\cmidrule(lr){2-4} \cmidrule(lr){5-6} \cmidrule(lr){7-8}
& {Coef} & {$t$} & {$R^2$}
& {Coef} & {$t$}
& {Coef} & {$t$} \\
\midrule
$\sigma$      & -3.611 & -0.80 & 0.27\% & -7.889 & -4.50 & -12.793 & -6.37 \\
Roll spread   &  0.083 &  7.60 & 4.59\% &        &       &   0.100 &  3.12 \\
LOGTRADESIZE  &  0.276 &  4.39 & 6.43\% &  0.228 &  2.80 &   0.036 &  0.43 \\
LOGNTRADE     &  0.016 &  0.92 & 0.20\% & -0.016 & -1.19 &  -0.035 & -2.54 \\
MONTHVOL      &  0.276 &  0.83 & 0.21\% &  0.652 &  2.49 &   0.584 &  3.23 \\
LOGP          & -0.129 & -2.77 & 5.20\% & -0.118 & -2.35 &  -0.146 & -2.89 \\
ILLIQ         & -0.947 & -0.69 & 0.05\% & -0.059 & -0.06 &  -0.441 & -0.49 \\
\midrule
Constant      &        &       &      & -0.456 & -0.69 &   1.086 &  1.45 \\
$R^2$ (\%)    &        &       &      & 10.89  &       &  13.38  &       \\
Adj.\ $R^2$ (\%) &     &       &      & 10.77  &       &  13.24  &       \\
$N$           &        &       &      & {4,559} &      & {4,559} &       \\
\bottomrule
\end{tabular}
\begin{minipage}{0.95\textwidth}
\tiny
\textit{Notes:} The dependent variable is $ROUGH_{i,d}=1\{T_{i,d}\leq -1.645\}$, where $T_{i,d}$ is the roughness test statistic and the 5\% critical value is based on the half-normal with point mass. Regressors enter in raw units. Individual columns report separate linear probability models. Model (1) are separate simple linear regressions for each regressor.  Model (2) is a multivariate regression, which includes all available regressors excluding the Roll spread. Model (3) adds Roll spread to the joint specification. Standard errors are clustered by ticker. 
\end{minipage}
\end{table}

The results are reported in \Cref{tab:rough_dummy_liquidity_reversal}. The dominant empirical pattern is the central role of short-run price reversals as a predictor of rough noise. The standard liquidity proxies studied by \citet{Ait2009}, which they find to be closely related to the magnitude of microstructure noise, play a more nuanced role in our setting, a contrast we return to at the end of the section.

The strongest evidence comes from Roll spread. In the individual regression, it is positive and highly significant ($t = 7.60$) and explains $4.59\%$ of the variation. In the joint specification it remains positive and significant ($t = 3.12$), adding about $2.5$ percentage points of explained variation on top of the other regressors. Since Roll spread is a direct function of the negative first-order autocovariance of intraday price changes, this result points to short-run price reversals as the main empirical driver of rough noise, matching the mechanism emphasized by the model.

A closely related pattern emerges for $LOGTRADESIZE$. In the individual regression, average trade size is in fact the single strongest predictor of rough noise, with an $R^2$ of $6.43\%$, higher than that of Roll spread itself. Once Roll spread is included in the joint specification, however, the coefficient on $LOGTRADESIZE$ collapses from $0.23$ to $0.04$ and becomes insignificant. The interpretation is a mediation pattern. Larger trades consume more liquidity and generate larger temporary price impact \citep{Kyle1985, Hasbrouck1991}, which mechanically appears as negative autocovariance in observed returns. Conditioning on Roll spread, the direct measure of these reversals, the explanatory content of trade size is exhausted. The fact that no independent trade-size effect remains is informative. It suggests that the relevant trade-size channel for rough noise operates through \emph{temporary} impact (reversals) rather than through \emph{permanent} impact (information), since the latter would not be captured by Roll spread.

Intraday volatility and lagged monthly volatility enter the joint specification with opposite signs. $\sigma$ is strongly negative ($t = -6.37$) while $MONTHVOL$ is positive and significant ($t = 3.23$). Rough noise is therefore most likely on days with low current intraday volatility within stocks that have been historically volatile.

The coefficient on $LOGP$ is negative and significant ($t = -2.89$ in the joint with Roll spread). The negative coefficient is consistent with rough noise being driven by reversals that are \emph{large relative to the price level}, not by reversals in absolute terms.

The standard liquidity proxies $ILLIQ$ and $LOGNTRADE$ have no significant individual effect on rough-noise detection ($t = -0.69$ and $t = 0.92$). Neither variable directly captures the magnitude or frequency of short-run reversals, trading frequency in particular is necessary but not sufficient for reversals. 

This pattern of results helps explain why the relation between liquidity and rough noise is less direct than the relation between liquidity and noise magnitude documented by \citet{Ait2009}. Illiquidity may widen spreads, increasing the size of each reversal, but it may also reduce trading frequency and therefore reduce the number of reversals. The two effects can offset each other, leaving little net association between standard illiquidity measures and rough-noise detection. The empirical evidence here suggests that rough noise is best understood as a reversal-driven phenomenon, most prevalent on days when trading generates many large reversals relative to fundamental price movements, regardless of whether the underlying stock is liquid or illiquid in the broader sense.

\section{Conclusion}\label{sec:conclusion}

In this paper, we argue that microstructure noise can be understood as mechanisms that induce price reversals and negative autocorrelation in high-frequency returns. With this interpretation, a trawl process is a natural choice for a parametric model of microstructure noise. Motivated by this, we work with a microstructural model for tick-by-tick price changes that explicitly separates permanent price changes, modeled by an integer-valued Lévy process, from fleeting price changes due to noise, modeled by a trawl process. We show that this model converges to a standard semimartingale model for the permanent price process plus a rough noise term originating from fleeting price changes at the macroscopic scale. In this sense, we provide a microstructural foundation for the rough noise model suggested by \cite{ChongEtAl2021}, based on realistic tick-by-tick dynamics. We further derive an explicit trawl function for the microstructural model, which gives rise to a Gaussian moving average with a gamma kernel as the limiting model for the noise.

We derive an estimation method for the tick-by-tick model based on the generalized method of moments and prove that the estimator is consistent. We further derive two limit theorems for the estimator: one where all parameters are in the interior of the parameter space, and one where the roughness parameter is allowed to lie on the boundary, corresponding to non-rough noise. Using these limit theorems, we construct a feasible test statistic for hypotheses on the roughness parameter, which in particular allows us to test whether the noise is rough.

Through simulation, we show that the model parameters can be estimated accurately in finite samples. A second simulation study shows that the proposed test has power against rough alternatives, and that its distribution under the null approaches the theoretical limiting distribution as the sample size increases.

Finally, we apply the methodology to tick-level transaction data from the TAQ database for the DJIA constituents in 2024. We first identify stock-day pairs with evidence of microstructure noise, since the model parameters are not identifiable in the absence of noise. About two-thirds of the sample exhibits evidence of noise, with one-fifth of the days with noise being classified as rough according to our test. The estimated roughness parameters on rough days are typically closer to zero than those found previously, a finding made possible by the use of daily data.

We then study which market characteristics are associated with roughness detection. Following the spirit of \cite{Ait2009}, we regress a roughness detection dummy on liquidity, volatility, trading-activity, and transaction-cost proxies. In contrast to their finding that less liquid stocks have larger microstructure noise, we do not find that roughness is explained by broad illiquidity alone. Instead, roughness appears to be associated with short-run price reversals: it is more likely when reversals are large relative to overall return variation. Hence, rough noise is better understood as a reversal-driven phenomenon than as a simple consequence of illiquidity.

Overall, the empirical results support the central mechanism of the paper: fleeting microstructure effects at the tick level can generate rough noise at the macroscopic scale. They also suggest that the appearance of rough noise depends on the balance between trading frequency, reversal size, and efficient-price variation. Understanding the market-design and economic forces behind this balance is a promising direction for future research.

\pagebreak

\section*{Data Availability Statement}
The data that support the findings of this study were obtained from the
New York Stock Exchange (NYSE) Trade and Quote (TAQ) database under licence.
Restrictions apply to the availability of these data, which were used under
licence for this study and are therefore not publicly available. The data
can be obtained directly from NYSE at
\url{https://www.nyse.com/market-data/historical/daily-taq}, subject to
their licensing conditions.

\section*{Generative AI Disclosure}
During the preparation of this manuscript, the authors used the large
language models Claude Opus 4.8 and Claude Fable 5 (Anthropic) to assist
with writing code for the production of figures, and to check grammar and
provide suggestions for the wording of the text, with the aim of improving
the clarity and presentation of the manuscript. All output from these
tools was reviewed and edited by the authors, who take full responsibility
for the content of the publication.

\pagebreak

\begin{singlespace}
{\small
\bibliographystyle{chicago}
\bibliography{mybib-v3}
}
\end{singlespace}
\end{document}